# Journal Impact Factor and Peer Review Thoroughness and Helpfulness: A Supervised Machine Learning Study

*Anna Severin MSc,[1, 2] Michaela Strinzel MSc,[3] Matthias Egger MD,[1,3,4] Tiago Barros PhD,[5] Alexander Sokolov BSc, [6] Julia Vilstrup Mouatt PhD,[6] Stefan Müller Phd [7]*

[1] Institute of Social and Preventive Medicine, University of Bern, 3012 Bern, Switzerland

[2] Graduate School for Health Sciences, University of Bern, Bern, Switzerland

[3] Swiss National Science Foundation, Bern, Switzerland

[4] Population Health Sciences, Bristol Medical School, University of Bristol, Bristol, UK

[5] Faculty Opinions, London, UK

[6] Clarivate, London, UK

[7] School of Politics and International Relations, University College Dublin, Dublin, Ireland

Corresponding Author:

Professor Matthias Egger, MD MSc FFPH
University of Bern
Institute of Social and Preventive Medicine
Mittelstrasse 43
CH-3012 Bern, Switzerland
matthias.egger@ispm.unibe.ch

Word counts: main text 3499 words, abstract 314 words, box on what this study adds 149 words, 41 references, one file with supplementary materials.

## ABSTRACT

**Objectives:** To examine the thoroughness and helpfulness of peer review reports submitted to journals with different impact factors.

**Design:** Text analysis using trained machine learning models.

**Data source:** Random sample of 10,000 peer review reports submitted to medical and life sciences journals, drawn from Publons' database and stratified by journal impact factor.

**Main outcome measures:** Thoroughness of the review, indexed by sentences addressing content categories materials and methods, presentation and reporting, results and discussion, importance and relevance. Helpfulness, gauged by sentences providing suggestions and solutions, examples, praise, or criticism. Association between the prevalence of different content with the journal impact factor, across ten groups defined by journal impact factor deciles.

**Results:** A total of 1,644 journals were represented; 187,240 sentences were analyzed. The median journal impact factor ranged from 1.23 to 8.03 across the ten groups (lowest journal impact factor 0.21, highest 74.70). Sentences on materials and methods were more common in the highest journal impact factor journals than in the lowest impact factor group (difference 7.8 percentage points; 95% CI 4.9 to 10.7%). The trend for presentation and reporting went in the opposite direction, with the journals with the highest impact factors giving less emphasis to such content (difference -8.9%; 95% CI -11.3 to -6.5%). For helpfulness, reviews for higher impact factor journals devoted less attention to suggestions and solutions and provided fewer examples than lower impact factor journals. Few differences were evident for other content categories. The proportion of sentences allocated to different content categories varied widely, even within journal impact factor groups.

**Conclusions:** Peer review in journals with higher journal impact factor tends to be more thorough in discussing the materials and methods used but less helpful in terms of suggesting solutions and providing examples. Differences were modest and variability high, indicating that the journal impact factor is a bad predictor for the quality of peer review of an individual manuscript.

## WHAT IS ALREADY KNOWN ON THIS TOPIC

- The inappropriate use of the journal impact factor for assessing the quality of research or the impact of researchers is well documented.
- Surveys of authors show that the metric continues to be an important consideration in authors' choice of journals.
- Few studies assessed the quality of peer review, using checklists or text analysis, but none examined associations between peer review content and journal impact factor.

## WHAT THIS STUDY ADDS

- This study found that the length of reports and the attention paid to materials and methods increased with the journal impact factor while journals with lower journal impact factor paid more attention to the presentation and reporting of the work.
- The content of peer review reports varied widely between journals with similar impact factors; the journal impact factor is, therefore, a bad predictor for the quality of peer review of an individual manuscript.

## INTRODUCTION

Peer review is a process of scientific appraisal by which manuscripts submitted for publication in journals are evaluated by experts in the field for originality, rigour and validity of methods and potential impact.[1] Peer review is an important scientific contribution and is increasingly visible on databases and researcher profiles.[2,3] Thorough peer review is particularly critical in the medical sciences, where practitioners rely on clinical research evidence to make a diagnosis, prognosis and choose a therapy. Recent developments, such as the retraction of peer-reviewed COVID-19 publications in prominent medical journals[4] or the emergence of predatory journals,[5,6] have prompted concerns about the rigour and effectiveness of peer review. Despite these concerns, research into the quality of peer review is scarce. Little is known about the determinants and characteristics of high-quality peer review. The confidential nature of many peer review reports and the lack of databases and tools for assessing their quality have hampered larger-scale research on peer review.

In the absence of evidence on the quality of peer review implemented in a journal, proxy measures like the journal impact factor[7] are used to assess the quality of journals and, by extension, the quality of peer review. The journal impact factor was originally developed to help libraries make indexing and purchasing decisions for their collections. It is a journal-based metric calculated by dividing the number of citations in a given year for papers published in the previous two years by the number of articles published these two years.[7] The prestige of the journal impact factor and its association with academic promotion, hiring decisions, and research funding allocation have led scholars to seek publication in journals with high impact factors.[8]

Despite using the journal impact factor as a proxy for the quality of the journal, it is unclear how the peer review characteristics for that journal relate to this metric. We combined human coding of peer review reports and quantitative text analysis to examine the association between peer review characteristics and journal impact factor in the medical and life sciences, based on a large sample of peer review reports.

## METHODS

Our study was based on peer review reports submitted to Publons from January 24, 2014, to May 23, 2022. Publons (now part of Web of Science) is a platform for scholars to track their peer review activities and receive recognition for reviewing.[9] We hand-coded 2,000 sentences from a training set of peer review reports and categorised content related to thoroughness and helpfulness. We then trained supervised machine learning models to classify the sentences in peer review reports as contributing or not to any categories. After assessing face validity, we examined the association between the journal impact factor and the prevalence of relevant sentences in peer review reports submitted to medical and life sciences journals. We analysed the data in regression models accounting for the hierarchical nature of the data.

### Data sources

As of May 2022, the Publons database contained information on 15 million reviews performed and submitted by more than 1,150,000 scholars for approximately 55,000 journals and conference proceedings. Reviews can be submitted to Publons in different ways. When scholars review for journals partnering with Publons and wish recognition, Publons receives the review and some meta-data directly from the journal. Scholars can submit their reviews for other journals by either forwarding the review confirmation emails from the journals to Publons or by sending a screenshot of the review from the peer review submission system. Publons audits a random subsample of emails and screenshots by contacting editors or journal administrators.

We randomly selected the peer review reports for the training from a broad spectrum of journals covering all Clarivate's Essential Science Indicator (ESI) fields[10] except Physics, Space Science and Mathematics. Reviews from the latter fields contained many mathematical formulae, which were difficult to categorise. In the next step, we selected a stratified random sample of 10,000 verified pre-publication reviews for analysis. First, we limited the Publons database to reviews from medical and life sciences journals based on ESI research fields, resulting in a data set of approximately 5.2 million reviews. The ESI field Multidisciplinary was excluded as these journals publish articles not within the medical and life sciences field (e.g *PloS ONE*, *Nature*, *Science*). Second, we divided these reviews into ten equal groups based on journal impact factor deciles. Third, we randomly sampled 1,000 reviews from each of the ten groups. We excluded second-round peer review reports whenever this information was available. Second-round peer review reports are often shorter and less likely to include comments that fall within our content categories. We also retrieved the continent of the reviewer's institutional affiliation, the total number of publications of the reviewer, the start and end year of the reviewers' publications and gender, based on the *gender-guesser* Python package v0.4.0.

### Classification and validation

We trained two reviewers (ASE, MS) in coding sentences. After piloting and refining coding and establishing intercoder reliability, the reviewers labelled 2,000 sentences (1,000 sentences each). Based on the pilot data, we calculated Krippendorff's α, a measure of reliability in content analysis.[11] The coders allocated sentences to none, one, or several of the eight content categories. We selected categories based on prior work, including the Review Quality Instrument and other scales and checklists,[12] and previous studies using text analysis or machine learning to assess student and peer review reports.[13–17]

Our categories describe, first, the *Thoroughness* of a review, measuring the degree to which a reviewer comments on (1) *Materials and Methods* (Did the reviewer discuss the methods of the manuscript?), (2) *Presentation and Reporting* (Did the reviewer comment on the presentation and reporting of the paper?), (3) *Results and Discussion* (Did the reviewer comment on the results and their interpretation?), and (4) the paper's *Importance and Relevance* (Did the reviewer discuss the importance or relevance of the manuscript?). Second, we examined the *Helpfulness* of a review, based on comments on (5) *Suggestion and Solution* (Did the reviewer provide suggestions for improvement or solutions?), (6) *Examples* (Did the reviewer give examples to substantiate his or her comments?), (7) *Praise* (Did the reviewer identify strengths?), and (8) *Criticism* (Did the reviewer identify specific problems). Categories were rated on a binary scale (1 for yes, 0 for no). A sentence could be coded as 1 for multiple categories. The appendix gives further details (p 7-9). We used a Naïve Bayes algorithm to train the classifier and predict the absence or presence of the eight characteristics in each sentence of the peer review report.[18]

For validation, we first performed out-of-sample predictions for the eight content categories by running five-fold cross-validation on the hand-coded sample of 2,000 sentences. We divided the sample into five equally sized subsets and ran the cross-validation five times. In each run, we used 1,600 sentences in the training set to predict the quality indicators in the remaining 400 sentences. We calculated performance measures, including precision (i.e. the positive predictive value), recall (i.e. sensitivity), and the F1 score. The F1 score is a weighted mean of precision and recall and an overall measure of accuracy.[19] Second, we compared the percentage of sentences addressing each category between the human annotation dataset and the output from the machine learning model. Finally, we identified the unique words in each quality category in "keyness" analyses.[20] The unique words retrieved from the keyness analyses reflect typical words used in each content category.

### Statistical analysis

We used a series of linear mixed-effects models to examine the association between peer review characteristics and journal impact factor groups to account for the clustered and hierarchical nature of the data.[21] We added random intercepts for reviewer and journal to account for the data structure. The dependent variable was the percentage of sentences in a review allocated to one of the eight review content categories. The independent variable was the journal impact factor group. We controlled for the length of reviews since longer texts may have a higher probability of addressing more categories. In a sensitivity analysis we additionally controlled for discipline, the academic age (estimates as the period between first and most recent publication), and the number of reviews submitted by a reviewer. Finally we examined whether results were influenced by the reviewer's gender. We report the coefficients from the regression models with 95% confidence intervals (CI), which indicate the percentage point change of prevalence for a given journal impact factor group relative to the lowest group. All analyses were done in R (version 4.2.1, R Core Team, Vienna, Austria). The packages used for data preparation, text analysis, supervised classification, and regression models, were *quanteda*,[22] *lme4*,[23] and *tidyverse*.[24]

# RESULTS

The training of coders resulted in acceptable to good between-coder agreement, with an average Krippendorff's $\alpha$ across the eight categories of 0.70. The final analyses included 10,000 review reports with 187,240 sentences that 9,259 reviewers submitted for 9,590 unique manuscripts to 1,644 journals.

## Characteristics of the study sample

The sample of 10,000 included 5,067 reviews from the ESI research field of Clinical Medicine, 943 from Environment and Ecology, 942 from Biology and Biochemistry, 733 from Psychiatry and Psychology, 633 from Pharmacology and Toxicology, 576 from Neuroscience and Behaviour, 566 from Molecular Biology and Genetics, 315 from Immunology, and 225 from Microbiology.

Across the ten groups of journals defined by journal impact factor deciles (1=lowest, 10=highest), the median journal impact factor ranged from 1.23 to 8.03, the minimum ranged from 0.21 to 6.51 and the maximum from 1.45 to 74.70 (Table 1). The proportion of reviewers from Asia, Africa, South America and Australia/Oceania declined when moving from journal impact factor group 1 to group 10. In contrast, there was a trend in the opposite direction for Europe and North America. Information on the continent of affiliation was missing for 43.5% of reviews (4,355). The median length of peer review reports increased by about 202 words from group 1 (median number of words 185) to group 10 (387). Table 2 lists the ten journals from each journal impact factor group that provided the largest number of peer review reports. The appendix (p 1-7) gives the complete list of journals.

## Performance of classifiers

In the training dataset, the most common categories based on human coding were *Materials and Methods* (coded in 823 sentences or 41.2% out of 2,000 sentences), *Suggestion and Solution* (638 sentences; 34.2%) and *Presentation and Reporting* (626 sentences; 31.3%). In contrast, *Praise* (210; 10.5%) and *Importance and Relevance* (175; 8.8%) were the least common. On average, the training set had 444 sentences per category. In cross-validation, precision, recall and F1 scores (binary averages across both classes [absent/present]) were similar within categories (see appendix p 12). The classification was most accurate for *Example* and *Materials and Methods* (F1 score 0.71) and least accurate for *Criticism* (0.57) and *Results and Discussion* (0.61). The prevalence predicted from the machine learning model was generally close to the human coding: point estimates did not differ by more than 2 to 5 percentage points, and the confidence intervals of both measures overlapped. The exception was *Suggestion and Solution,* where the difference between the predicted prevalence and human coding was nine percentage points. Further details are given in the appendix (p 10-14).

## Content categories

The prevalence of sentences addressing each of the eight content categories are shown in Figure 1. The majority of sentences (114,710 sentences, 60.9%) contributed to more than one content category; a minority (18,244 sentences, 9.69%) were not assigned to any category. The content categories *Suggestion and Solution*, *Materials and Methods*, and *Presentation and Reporting* of manuscripts were most extensively covered. On average, a third or more of the sentences in a peer review report addressed these categories. In contrast, only 15,366 sentences (9.20%) addressed *Importance and Relevance* of the study. *Criticism* was more common than *Praise* (32,665 sentences, 16.1% vs 18,057 sentences, 13.5%). Most distributions were skewed right, with a peak at 0% showing the number of reviews that did not address the content category (Figure 1).

Figure 2 shows the estimated prevalence of sentences addressing the eight content categories across the ten journal impact factor groups. Among thoroughness categories, the percentage of sentences addressing *Materials and Methods* increased from 41.5% to 52.0% from journal impact factor group 1 to 10. In contrast, the focus on *Presentation and Reporting* declined with increasing JIF (from 35.8% to 27.8%). The attention given to *Results and Discussion* increased slightly, whereas the attention to *Importance and Relevance* declined with increasing Journal Impact Factor groups from 11.3% to 8.9%. For helpfulness, the percentage of sentences including *Suggestion and Solution* declined, from 41.9% to 34.3% with increasing journal impact factor. No clear trends were observed for *Example*, *Praise*, and *Criticism*. The distribution of prevalences was broad, even within groups of journals with similar impact factors.

## Regression analysis

The percentage point changes in sentences addressing content categories by journal impact factor group estimated from the linear mixed-effects models are shown in Figure 3. All models control for review length, and include random intercepts for the the journal name and reviewer ID. The coefficients and standard errors are available in the online appendix (p 16-17). The results confirm those observed in the descriptive analyses. First, for thoroughness, the prevalence of sentences on *Materials and Methods* in the journals with the highest impact factor was higher (+7.8 percentage points; 95% CI +4.9 to +10.7) than in the group with the lowest impact factor journals. The trend for sentences addressing *Presentation and Reporting* went in the opposite direction, with reviews submitted to the journals with the highest impact factor giving less emphasis to such content (-8.92 percentage points; 95% CI -11.3 to -6.5). A small difference in the same direction was observed for sentences addressing *Importance and Relevance* (-2.0; 95% CI -3.3 to -0.7) whereas no difference was evident for *Results and Discussion*.

Second, for sentences reflecting helpfulness, reviews for higher impact factor journals devoted less attention to *Suggestions and Solutions* and provided fewer *Examples* than lower impact factor journals. The group with the highest journal impact factor had 8.5 percentage points fewer sentences addressing *Suggestion and Solution* (95% CI -10.8 to -6.2) and fewer sentences providing *Examples* (-2.6 percentage points; 95% CI -4.4 to -0.8). No differences were observed for *Praise* and *Criticism* (Figure 3). Associations were approximately linear across the categories. The sensitivity analysis showed that adjusting for additional variables (discipline, career stage of reviewers, and logged number of reviews submitted) tended to strengthen relationships between content categories and journal impact factor. Results were generally similar for male and female reviewers, and when replacing the journal impact factor groups with the raw journal impact factor (appendix, p 18-22).

## Typical words in content categories

The keyness analyses of the set of 2,000 sentences showed that typical and unique words in the thoroughness categories were 'data', 'analysis', 'method' (*Materials and Methods*); 'text', 'figure', 'sentence' (*Presentation and Reporting*); 'results', 'discussion', 'findings' (*Results and Discussion*); and 'contribution', 'literature', 'topic' (*Importance and Relevance*). For helpfulness, common unique words included 'please', 'need', 'include' (*Suggestion and* Solution); 'line', 'page', 'figure' (Examples); 'interesting', 'good', 'well' (Praise), and 'however', '(un)clear', 'mistakes' (Criticism). The appendix (p 15) provides further details.

## DISCUSSION

This study used supervised machine learning to analyse the content of peer review reports and investigate the association of content with the journal impact factor. We found that the impact factor was associated with the characteristics and content of peer review reports and reviewers. The length of reports and the attention paid to materials and methods increased with increasing journal impact factor. Conversely, the prevalence of sentences including suggestions and solutions, examples or addressing the presentation and reporting of the work declined with increasing journal impact factor. Finally, the proportion of reviewers from Asia, Africa and South America declined with increasing journal impact factor, whereas the proportion of reviewers from Europe and North America increased.

The limitations of the journal impact factor are well documented[25–27] and there is increasing agreement that its should not be used to evaluate the quality of research published in a given journal. The San Francisco Declaration on Research Assessment (DORA) calls for the elimination of any journal-based metrics in funding, appointment, and promotion.[28] DORA is supported by thousands of universities, research institutes and individuals. Our study shows that the peer reviews submitted to journals with higher journal impact factor may be more thorough than those submitted to lower impact journals. Should, therefore, the journal impact factor be rehabilitated, and used as a proxy measure for peer review quality? Similar to the distribution of citations in a journal, the prevalence of content related to thoroughness and helpfulness varied widely even between journals with similar journal impact factor. In other words, the journal impact factor is a poor proxy measure for the thoroughness or helpfulness of peer review authors may expect when submitting their manuscript. Rather, journals and funders could use our approach to analyse the thoroughness and helpfulness of their peer review. Journals could submit all their peer review reports to an independent organisation for analysis. The results could help journals improve peer review, inform the training of peer reviewers, and help readers gauge the quality of the journals in their field.

The higher proportion of reviewers from Europe and North America and the fact that reviewers with English as their first language tend to write longer reports could partly explain the increase in the length of reports with increasing journal impact factor. Further, high journal impact factor journals may be more prestigious to review for and can thus afford to recruit more senior scholars. More senior researchers might be more likely to focus on arguably the most important aspects of a paper, such as the methodology, and these journals may put more emphasis on methodological aspects. Junior reviewers might be less able to comment on the methodology and might focus on more superficial aspects of a manuscript such as grammar issues, typos, and commenting on the presentation and reporting. Of note, there is evidence to the opposite, suggesting that the quality of reports decreases with age or years of reviewing.[30,31] Interestingly, several medical journals with high impact factors have recently committed to improving diversity among their reviewers.[32–34] Unfortunately, due to incomplete data, we could not examine the importance of the level of seniority of reviewers. Adjusting analyses for the length of peer reviews did not change associations between content and journal impact factor. Therefore, longer peer reviews may not necessarily address more aspects of the study under review.

Peer review reports have been hidden for many years, hampering research on their characteristics. Previous studies were based on smaller, selected samples. An early trial evaluating the effect of blinding reviewers to the authors' identity on the quality of peer review was based on 221 reports submitted to a single journal.[35] Since then, science has become more open, embracing open access to publications and data, and open peer review. Some journals now publish peer reviews and authors' responses with the articles.[36,37] Bibliographic databases have also started to publish reviews.[38] The European Cooperation in Science and Technology (COST) Action on new frontiers of peer review (PEERE), established in 2017 to examine peer review in different areas, was based on data from several hundred Elsevier journals from a wide range of disciplines.[39]

To our knowledge, the Publons database is the largest of peer review reports, and the only one not limited to individual publishers or journals, making it a unique resource for research on peer review. Based on 10,000 peer review reports submitted to medical and life science journals, this is likely the largest study of peer review content ever done. It built on a previous analysis of the characteristics of scholars who review for predatory and legitimate journals.[40] Other strengths of this study include the careful classification and validation step, based on the coding by hand of 2,000 sentences by trained coders. Our study also has weaknesses. Reviewers may be more likely to submit their review to Publons if they feel it meets general quality criteria. This could have introduced bias if the selection process into Publons' database depended on the journal impact factor. However, the large number of journals within each journal impact factor group makes it likely that the patterns observed are real and generalisable. Also, we trained the algorithm on journals from many disciplines, which should make it applicable to other fields than medicine and the life sciences. Our study would have benefited from further increasing the training set size and using transformer-based machine learning models.[41] We acknowledge that our findings are more reliable for the more common content categories than for the less common. Finally, we could not assess to what extent the content of peer review reports affected acceptance or rejection of the paper.

In conclusion, this study of peer review characteristics indicates that peer review in journals with higher impact factors tends to be more thorough in addressing study methods but less helpful in suggesting solutions or providing examples. However, differences were modest, and the journal impact factor is a bad predictor of the quality of peer review of an individual manuscript.

## Contributors

ASE, TB, ME and SM conceptualised and designed the study. ASO prepared the study data set. ASE developed the codebook and ASE and MS coded the sentences. SM performed statistical and machine learning analyses. ASE drafted the first version of the manuscript, which was further developed by ME, SM, MS, JVM, TB and ASO. SM and ME accessed and verified the aggregated analysis dataset and take responsibility for the integrity of the data and the accuracy of the data analysis. All authors contributed to data interpretation, critical review, and manuscript revision. The corresponding author attests that all listed authors meet authorship criteria and that no others meeting the criteria have been omitted.

## Funding

Swiss National Science Foundation (SNSF, grant 189498). The views expressed are those of the authors and not necessarily those of the SNSF.

## Ethical approval

Not required.

## Competing interests

All authors have completed the ICMJE uniform disclosure form at www.icmje.org/disclosure-of-interest/. Two of the authors (MS, ME) were employed by the SNSF and one (ASE) was a PhD student supported by the SNSF at the time of the study. One author (ASE) was employed by Accenture at some time of the study. Three authors (TB, ASO, JVM) were employed by Publons (now a part of Web of Science).

## Data sharing

While the review texts cannot be shared, and the identifiers of journals and authors are anonymous, the aggregated and classified dataset will be shared on Harvard Dataverse to ensure the computational reproducibility of all findings of this paper.

## Dissemination to participants and related patient and public communities

Not applicable.

## Acknowledgements

We are grateful to Anne Jorstad from the data team of the Swiss National Science Foundation (SNSF) for helpful comments on an earlier draft of this paper. We would also like to thank Marc Domingo (Publons, now part of Web of Science) for helpful feedback on the sampling procedure.

## Table 1: Characteristics of peer review reports by journal impact factor group.

| | Journal impact factor group | | | | | | | | | |
|---|---|---|---|---|---|---|---|---|---|---|
| | 1 | 2 | 3 | 4 | 5 | 6 | 7 | 8 | 9 | 10 |
| Median JIF (range) | 1.23 (0.21-1.45) | 1.68 (1.46-1.93) | 2.07 (1.93-2.22) | 2.42 (2.23-2.54) | 2.77 (2.54-3.01) | 3.26 (3.01-3.55) | 3.83 (3.55-4.20) | 4.53 (4.21-5.16) | 5.67 (5.163-6.5) | 8.03 (6.51-74.70) |
| No. of review reports | 1000 | 1000 | 1000 | 1000 | 1000 | 1000 | 1000 | 1000 | 1000 | 1000 |
| No. of journals | 256 | 224 | 151 | 146 | 183 | 156 | 155 | 129 | 98 | 146 |
| No. of reviewers | 967 | 960 | 969 | 958 | 965 | 973 | 961 | 939 | 970 | 962 |
| No. of sentences (median; IQR) | 9 (4-18) | 11 (6-22) | 12 (5-22) | 13 (6-23) | 14 (7-25) | 14 (7-25) | 16 (8-28) | 17 (8-27) | 16.5 (9-27) | 18 (10-30) |
| No. of words (median; IQR) | 185 (84-359) | 232.5 (116-426) | 225 (104-419) | 256.5 (116-478) | 284.5 (146-506) | 271 (142-495) | 346 (170-581) | 344.5 (176-555) | 350.5 (195-567) | 387 (213-672) |
| Contintent of reviewers' affiliation | | | | | | | | | | |
| Asia | 139 | 107 | 163 | 115 | 93 | 135 | 98 | 93 | 80 | 62 |
| Africa | 15 | 14 | 18 | 9 | 5 | 14 | 8 | 6 | 5 | |
| Europe | 119 | 156 | 187 | 190 | 231 | 250 | 268 | 273 | 280 | 241 |
| North America | 97 | 113 | 105 | 153 | 162 | 151 | 191 | 180 | 166 | 213 |
| Central/South America | 61 | 42 | 36 | 25 | 38 | 22 | 22 | 20 | 23 | 10 |
| Australia/Oceania | 50 | 55 | 36 | 46 | 64 | 37 | 26 | 37 | 38 | 52 |
| Missing | 519 | 513 | 455 | 462 | 407 | 391 | 387 | 391 | 408 | 422 |
| Gender of reviewer | | | | | | | | | | |
| Female | 242 | 262 | 261 | 254 | 241 | 211 | 216 | 189 | 260 | 206 |
| Male | 518 | 516 | 478 | 549 | 548 | 551 | 575 | 584 | 543 | 599 |
| Unknown | 240 | 222 | 261 | 197 | 211 | 238 | 209 | 227 | 197 | 195 |

IQR, interquartile range.
Contintents are ordered by population size.
Journal impact factor group defined by deciles (1=lowest, 10=highest).

**Table 2: The ten journals from each journal impact factor group that provided the largest number of peer review reports.**

| Journal impact factor group | | | | | | | | | |
|---|---|---|---|---|---|---|---|---|---|
| **1** | **2** | **3** | **4** | **5** | **6** | **7** | **8** | **9** | **10** |
| J of International Medical Research (31) | European J of Ophthalmology (35) | International J of Dermatology (58) | BMJ Open (103) | Physics in Medicine and Biology (64) | Environmental Science and Pollution Research (59) | Magnetic Resonance in Medicine (40) | Plastic and Reconstructive Surgery (65) | Bioinformatics (71) | Nucleic Acids Research (86) |
| Echocardiography (29) | J of Cardiac Surgery (32) | Frontiers in Psychology (45) | The Laryngoscope (60) | J of Advanced Nursing (36) | Cancer Medicine (48) | J of Magnetic Resonance Imaging (38) | Human Brain Mapping (51) | British J of Surgery (62) | American J of Transplantation (66) |
| International J of Environmental Analytical Chemistry (24) | J of Cosmetic Dermatology (26) | Human & Experimental Toxicology (43) | Dermatologic Therapy (51) | Transfusion (31) | J of Biomolecular Structure and Dynamics (37) | J of Neural Engineering (29) | Frontiers in Microbiology (35) | Molecular Ecology (60) | Allergy (62) |
| ANZ J of Surgery (23) | Clinical Transplantation (23) | J of Clinical Nursing (38) | International J of Clinical Practice (41) | International Forum of Allergy & Rhinology (25) | The J of Dermatology (29) | J of Thrombosis and Haemostasis (29) | International J of Cancer (33) | Diabetes, Obesity and Metabolism (57) | eLife (55) |
| J of Orthopaedic Surgery (23) | J of Clinical Laboratory Analysis (22) | Natural Product Research (34) | Pediatric Blood and Cancer (41) | Head & Neck (23) | Transplant International (28) | Phytotherapy Research (29) | Frontiers in Immunology (30) | Environmental Research Letters (55) | Ecology Letters (42) |
| J of Obstetrics and Gynaecology Research (19) | The J of Maternal-Fetal & Neonatal Medicine (22) | The British J of Radiology (29) | Pediatric Pulmonology (32) | Oral Diseases (23) | Oikos (26) | BMC Genomics (28) | Cancer Science (29) | J of Cellular Physiology (37) | Hepatology (38) |
| Pediatrics International (19) | Acta Radiologica (21) | Environmental Technology (26) | Artificial Organs (24) | J of Pharmacy and Pharmacology (22) | J of Gastroenterology and Hepatology (25) | British J of Clinical Pharmacology (27) | Transplantation (28) | Rheumatology (33) | Global Change Biology (35) |
| Pediatric Dermatology (17) | American J of Perinatology (20) | Technology in Cancer Research & Treatment (21) | J of Cardiovascular Electrophysiology (24) | Frontiers in Neurology (21) | European J of Neuroscience (24) | Applied and Environmental Microbiology (25) | Frontiers in Oncology (26) | J of Bone and Mineral Research (31) | British J of Dermatology (34) |
| International J of Ophthalmology (16) | Current Eye Research (15) | Andrologia (20) | Physiological Measurement (24) | Colorectal Disease (20) | Reproduction (24) | Diseases of the Colon & Rectum (25) | Antimicrobial Agents and Chemotherapy (24) | Liver International (24) | IEEE Transactions on Medical Imaging (30) |
| Pacing and Clinical Electrophysiology (16) | Australasian J of Dermatology (14) | Brain and Behavior (20) | Annals of Pharmacotherapy (20) | The J of Clinical Hypertension (18) | Scandinavian J of Medicine & Science in Sports (24) | J of Biogeography (23) | BJU International (24) | J of Antimicrobial Chemotherapy (23) | Alimentary Pharmacology & Therapeutics (27) |

The journal (J) and number of review reports (in brackets) are listed. Journal impact factor groups were defined by deciles (1=lowest, 10=highest). The complete list of the 1,664 journals is available from the online appendix (p 1-7).

**Figure 1: Distribution of sentences in peer review reports allocated to eight content categories.**

The percentage of sentences in a review allocated to the eight peer review content categories is shown. A sentence could be allocated to no, one, or several categories. Analysis based on 10,000 review reports. Vertical dashed lines show the average prevalence.

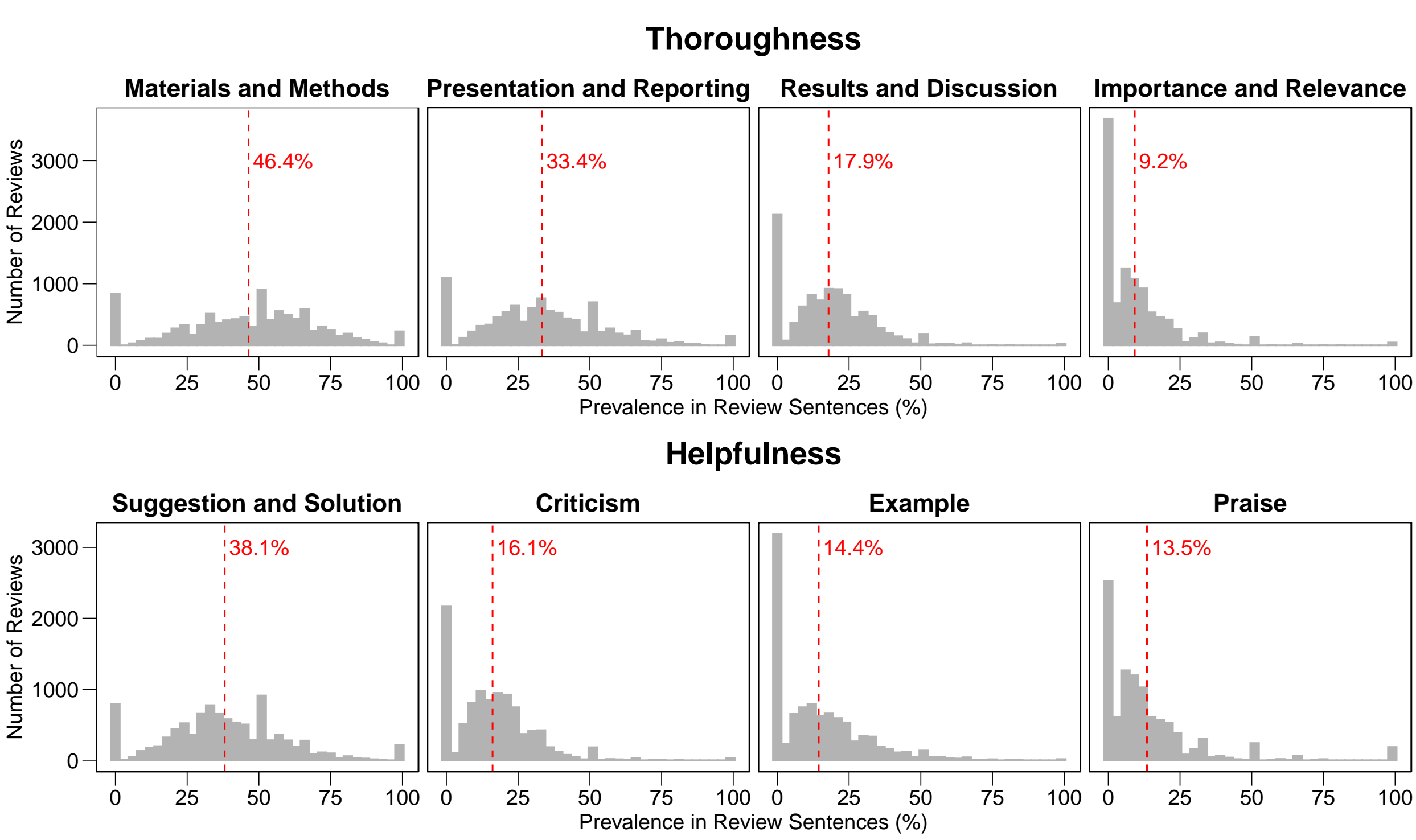

**Figure 2: Distribution of sentences in peer review reports allocated to eight content categories by journal impact factor group.**

The percentage of sentences in a review allocated to the eight peer review quality categories is shown. A sentence could be allocated to no, one, or several categories. Analysis based on 10,000 review reports. Vertical dashed lines show the average prevalence.

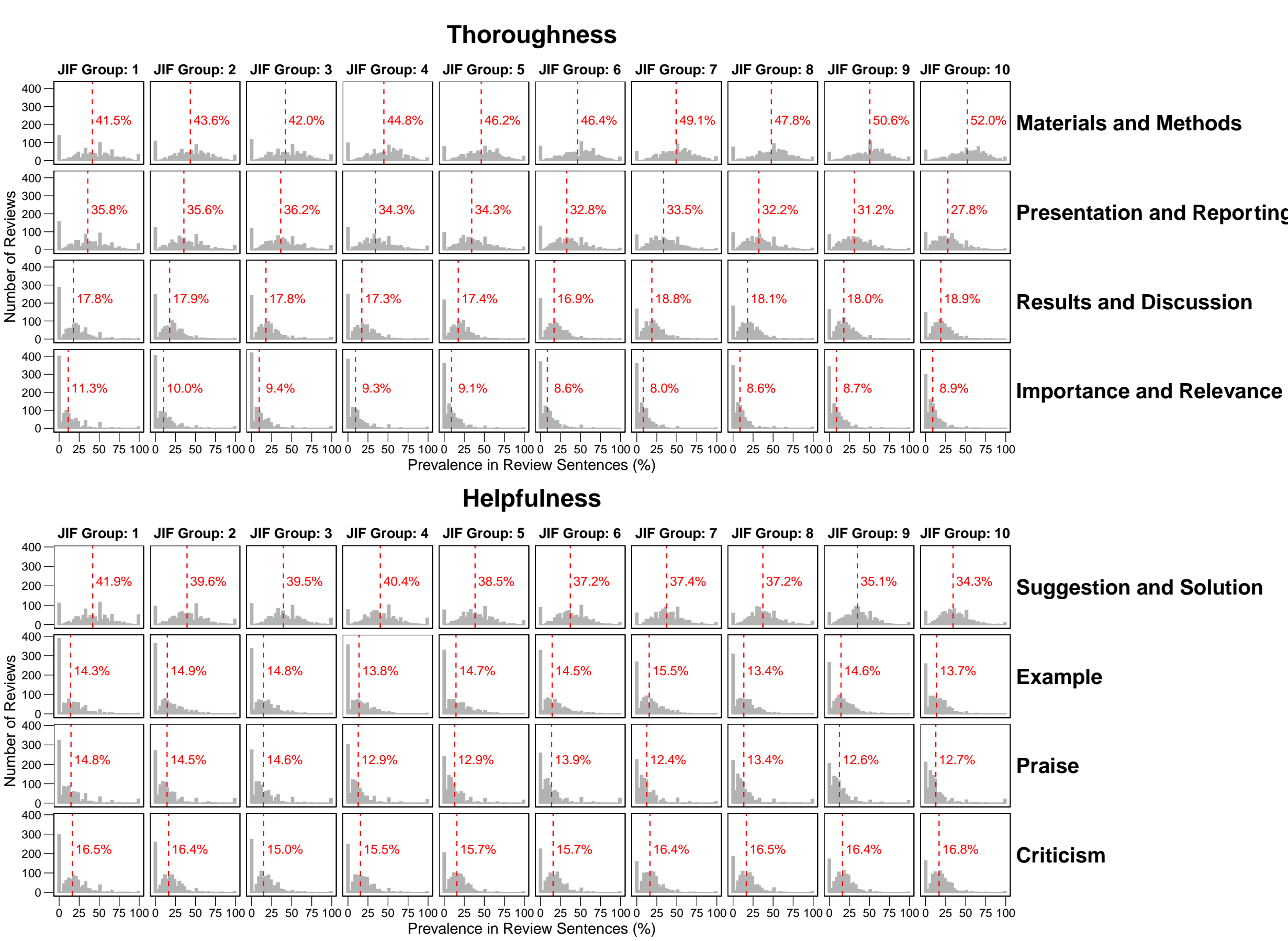

**Figure 3: Percentage point change in the proportion of sentences addressing thoroughness and helpfulness categories, relative to the lowest journal impact factor group.**

Regression coefficients and 95% confidence intervals are shown. Analysis based on 10,000 review reports. All linear mixed-effects models control for review length and include random intercepts for the journal ID and reviewer ID.

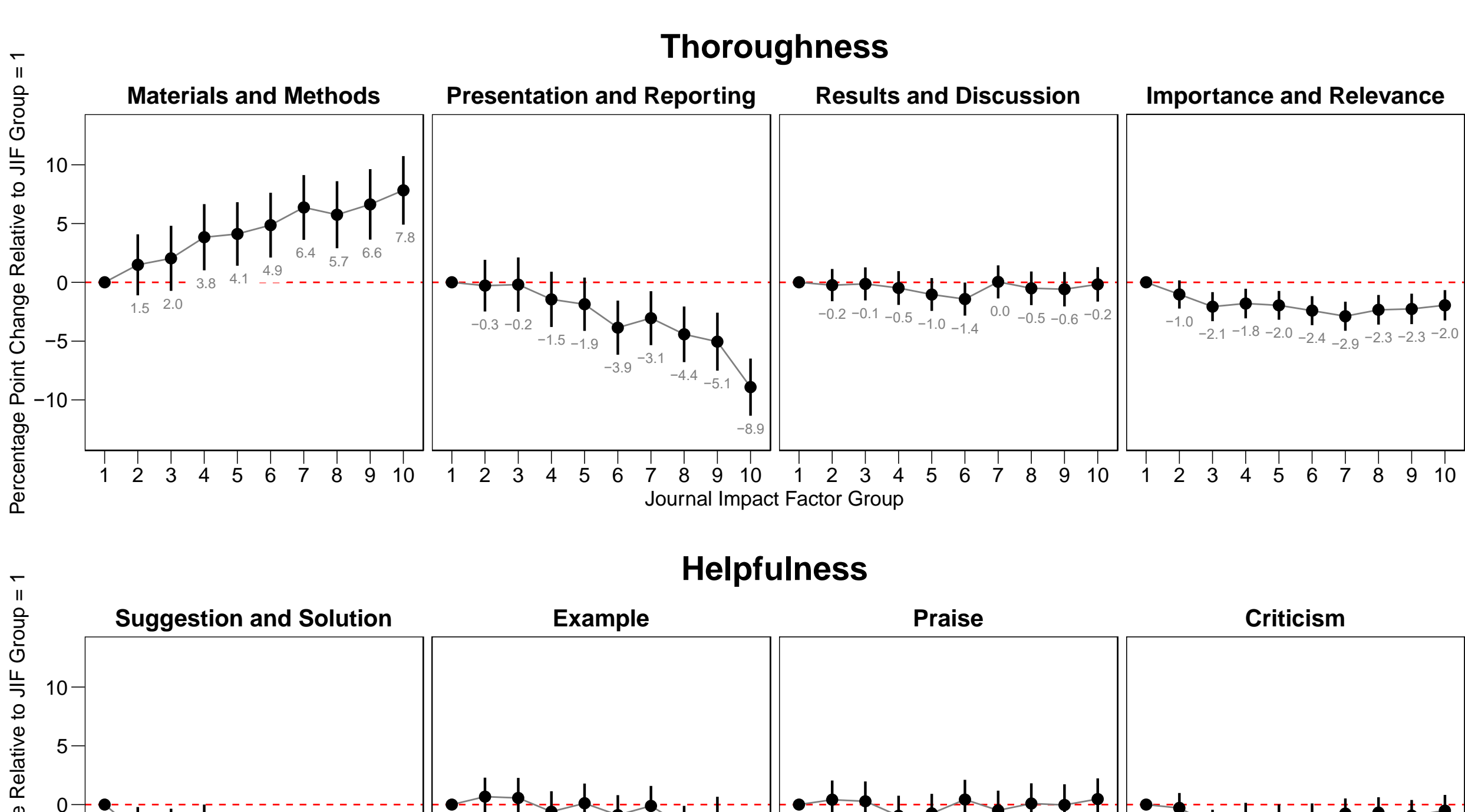